\documentclass[shortnote,twocolumn]{jpsj3}
\usepackage{txfonts}
\usepackage[version=4]{mhchem}
\usepackage{xspace}
\usepackage{float}
\usepackage[dvipdfmx]{graphicx}
\usepackage{color}

\newcommand{\CS}{$\mathrm{CaSb}_2$}
\newcommand{\Tc}{$T_{\mathrm{c}}$}

\title{Electronic state of superconductivity in line nodal material \CS\ under pressure up to $4.2 \,\mathrm{GPa}$}

\author{
    Shumpei Oguchi$^1$ \thanks{oguchi.shumpei.77r@st.kyoto-u.ac.jp}, Kenji Ishida$^1$, Atsutoshi Ikeda$^2$, Yoshiteru Maeno$^3$, and Shunsaku Kitagawa$^1$ \thanks{kitagawa.shunsaku.8u@kyoto-u.ac.jp} 
}

\inst{
    $^1$ Department of Physics, Kyoto University, Kyoto 606-8502, Japan\\ 
    $^2$Department of Electronic Science and Engineering, Kyoto University, Kyoto 615-8510, Japan \\ 
    $^3$Toyota Riken\,-\,Kyoto University Research Center, Kyoto University, Kyoto 606-8501, Japan
} 

\abst{
We report the results of resistance measurements under pressure up to $4.2\,\mathrm{GPa}$ on single-crystalline \CS\ , which shows the maximum of superconducting transition temperature \Tc\ at $3.1\,\mathrm{GPa}$. At room temperature, $R(P)$ shows a subtle anomaly at $3.1\,\mathrm{GPa}$. However, Bloch-Gr\"uneisen analysis of $R(T)$ indicates that the electronic state does not change significantly across $3.1\,\mathrm{GPa}$.
}
\begin{document}
\maketitle
\CS\ has a monoclinic structure with non-symmorphic space group (No.11, $P2_1/m$, $C_{2h}^2$)\cite{funada2019}. This non-symmorphic crystalline symmetry protects \CS\ as a line-nodal semimetal even with spin--orbit coupling(SOC)\cite{funada2019}. 
At ambient pressure, \CS\ shows superconductivity below the superconducting transition temperature \Tc $\, = 1.7 \,\mathrm{K}$\cite{ikeda2020}, considered as conventional weak-coupling $s$-wave type\cite{takahashi2021}. Under pressure, however, \Tc\ shows a non-monotonic peak behavior with maximum value \Tc\ $= 3.4 \,\mathrm{K}$ around $P_{\mathrm{max}} = 3.1 \,\mathrm{GPa}$\cite{kitagawa2021}.

According to the BCS theory\cite{bardeen1957}, the superconducting transition temperature \Tc\ is described as follows:
\begin{equation}
    T_c = \frac{1.13 \hbar \omega_\mathrm{D}}{k_\mathrm{B}} \exp \left( -\frac{1}{N(E_F;P)V}\right),
\end{equation}
where $N(E_\mathrm{F};P)$, $\omega_\mathrm{D}$, and $V$ are the density of states (DOS) at the Fermi level as a function of pressure: $P$, Debye frequency, and electron-phonon interaction, respectively.

A common explanation for change in \Tc\ under pressure is that the pressure dependence of $N(E_\mathrm{F};P)$ leads to a change in \Tc\ , especially in weak coupling superconductors\cite{smith1967}. However, nuclear quadrupole resonance (NQR) measurements under pressure have indicated that the $N(E_\mathrm{F};P)$ remains essentially unchanged by pressure at least up to $2.08\,\mathrm{GPa}$\cite{takahashi2024}.Therefore, the peak in \Tc\ is possibly related to pressure-induced changes in $V$ and $\omega_\mathrm{D}$, both lattice-related quantities. Indeed, X-ray diffraction (XRD) measurements under pressure suggest a first-order structural transition at $P_{\mathrm{max}}$ without symmetry change\cite{takahashi2025}.

Here, one possible scenario is that pressure induces a structural transition that predominantly modifies lattice-related quantities such as $\omega_\mathrm{D}$ and $V$ without a large change in the DOS, leading to the non-monotonic \Tc. In this note, we report the resistance measurements under pressure up to $4.2 \,\mathrm{GPa}$ to investigate a change in the phonon properties.

Single crystals of \CS\ were synthesized by the Sb self-flux method\cite{ikeda2022}. In this study, the sample1 and sample2 of \CS\ were taken from a large bulk crystal. Electrical resistance measurements on \CS\ and Pb were performed by the conventional four-terminal method. It is noted that the sample sizes are too small to estimate the resistivity. Four terminals of Cu wires were attached by a spot-welding method. Pressure was applied using an indenter-type cell\cite{kobayashi2007}, with Daphne 7575 as the pressure-transmitting medium. The pressure at room temperature was estimated from the pressure dependence of the resistance of Pb\cite{eiling1981}. During the temperature-dependent measurements, the pressure was determined using $P = \{T_c^{\mathrm{Pb}}(P=0) - T_c^{\mathrm{Pb}}(P)\}/{0.364},$ where $P$ is in GPa\cite{eiling1981}. 
\begin{figure}[tbp]
    \begin{center}
        \includegraphics[width=8cm]{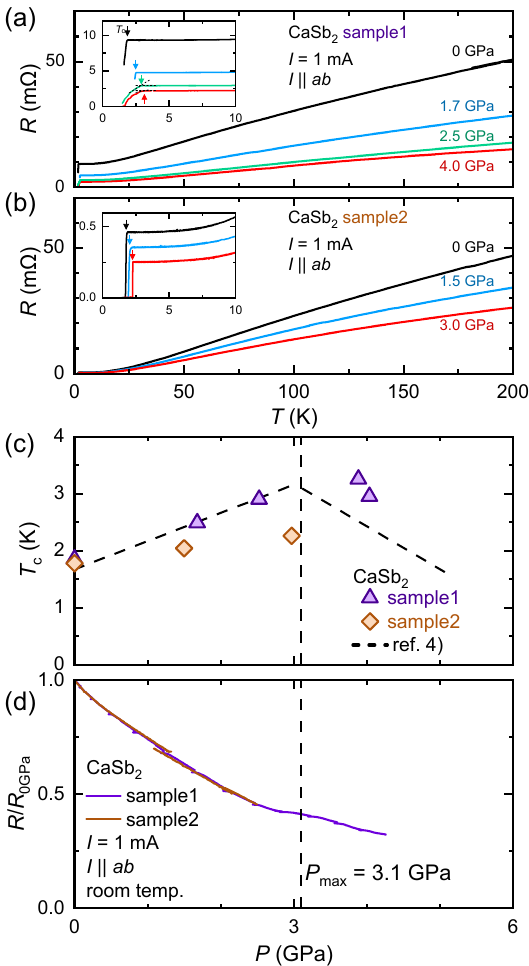}
    \end{center}
    \caption{(Color online)\, Temperature dependence of resistance under various pressures for (a) the sample1 and (b) the sample2. The arrows in the insets represent \Tc\ determined from the onset of superconductivity. (c) Pressure dependence of \Tc\ of \CS. The dashed lines are guides to the eye taken from ref. \cite{kitagawa2021}. (d) Pressure dependence of the resistance normalized at ambient pressure.} 
    \label{result}
\end{figure}
We performed electrical resistance measurements under pressure using single-crystals of \CS\ and reproduced the peak in \Tc\ as reported on polycrystals as shown in Fig.\ \ref{result}. We determined \Tc\ from the onset of the superconducting transition. As shown in Figs.\ \ref{result} (a) and \ref{result} (b), two samples show quite different residual resistance. The residual resistivity ratios (RRRs) given as $R(200\,\mathrm{K}) / R(4.2\,\mathrm{K})$ are 5.4 and 101 for the sample1 and sample2, respectively. Although sample 1 shows a slightly higher \Tc\ , its broader transition and much smaller RRR indicate lower quality, probably due to being taken from near the crystal surface. While the sample1 has lower quality, only the resistance of the sample1 could be measured under pressures exceeding $P_{\mathrm{max}}$. As discussed below, the behavior of quantities except RRR is almost the same. 

Figure \ref{result}(d) shows the pressure dependence of the resistance at room temperature. The sample2 has data only up to $2.5 \,\mathrm{GPa}$, but the curves overlap well with that of the sample1. Around $P_{\mathrm{max}}$, a subtle change in the slope of $R(P)$ was observed.
\begin{figure}[tbp]
    \begin{center}
        \includegraphics[width=8cm]{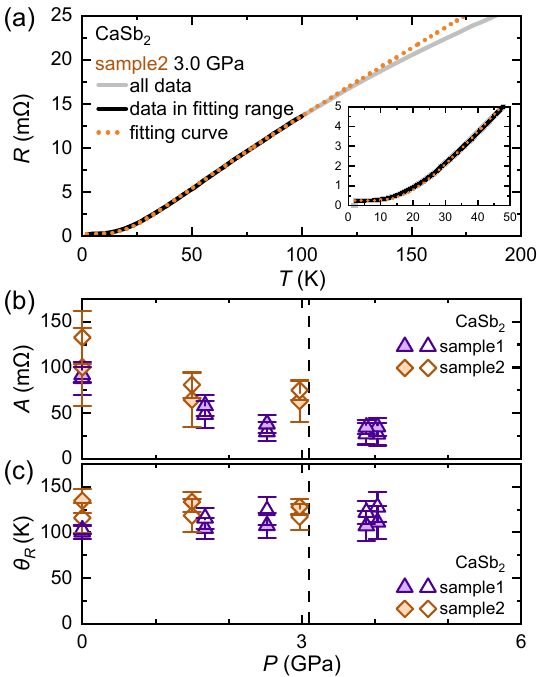}
    \end{center}
    \caption{(Color online)\, (a)\ Temperature dependence of resistance of the sample2 at $3.0 \,\mathrm{GPa}$. Solid line is experimental data, dashed line is fitting line with eq.(3) using the normal-state data up to $100 \,\mathrm{K}$. (b)(c)\ Pressure dependence of the fitting parameters, $A$, $\theta_R$. Closed symbols mean that fitting range is up to $50\,\mathrm{K}$. Open symbols mean up to $100\,\mathrm{K}$.}
    \label{fit}
\end{figure}
To investigate changes in the electronic state at low temperatures, we analyzed the temperature dependence of the resistance of the sample2 using the Bloch-Gr\"uneisen formula\cite{patterson2010}:
\begin{equation}
    R(T) = R_0 + A \left(\frac{T}{\theta_R}\right)^5 \int_{0}^{\theta_R/T} \frac{x^5}{(e^x - 1)(1 - e^{-x})} dx,
    \label{fitting}
\end{equation}
where $R_0$ is the residual resistance. The phonon contribution to the resistance is represented by the second term, in which $\theta_R$ is the Bloch-Gr\"uneisen temperature. Here, $A$ and $\theta_R$ are the fitting parameters. The fittings were performed in the normal state up to $50 \,\mathrm{K}$ and $100 \,\mathrm{K}$ to check whether the resulting fitting parameters are robust and do not strongly depend on the fitting range. The fitting example is shown in Fig.\ \ref{fit}(a), and the pressure dependence of the fitting parameters for each sample is shown in Figs.\ \ref{fit}(b) and (c). As seen in the figures, no significant pressure dependence was observed in either sample or in either fitting range. In the Bloch--Gr\"uneisen analysis, $\theta_R$ is the Debye temperature as obtained from resistance measurements, i.e., a quantity that can be an indicator of $\omega_\mathrm{D}$. The coefficient $A$ reflects the transport electron--phonon coupling and is an indicator of electron--phonon interaction $V$\cite{allen1987}. Although $A$ ($\theta_R$) are not identical to $V$($\omega_\mathrm{D}$) strictly, both $A$ and $\theta_R$ can qualitatively trace the changes by pressure. Therefore, these results suggest that the electronic state, especially, phonon properties does not significantly change at $P_{\mathrm{max}}$ within our resolution. It is interesting that the electronic state exhibits only subtle changes at $P_{\mathrm{max}}$, although the structural transition occurs as first-order and \Tc\ shows the peak at $P_{\mathrm{max}}$.

The origin of the slope change in $R(P)$ observed at $P_{\mathrm{max}}$ is considered to be a change in the response of the Sb(1) network to pressure reported by Takahashi \textit{et al.}\cite{takahashi2025}. Such a subtle change was not seen in polycrystals, likely due to grain-boundary scattering.\cite{kitagawa2021}
One possible scenario is whereas resistance reflects the average contribution of all phonons, \Tc\ is affected contribution from specific phonon modes.

In conclusion, we performed the electrical resistance measurements on single-crystal \CS\ under pressure. The peak structure of \Tc\ at $P_{\mathrm{max}}$ was successfully reproduced, whereas no appreciable change in the low-temperature electronic state was detected within our experimental resolution. This implies that the electronic state remains essentially unchanged up to the highest pressure region. To clarify origin of the peak in \Tc, further studies are required; in particular, high-pressure NQR would help constrain the superconducting symmetry and its coupling to lattice dynamics. In addition, since \CS\ has been discussed as a candidate topological material, the relationship between potential topological features and the observed enhancement of superconductivity remains an open question for future work.
\begin{acknowledgment}
This work was supported by Grants-in-Aid for Scientific Research (KAKENHI Grant Nos. JP20KK0061, JP20H00130, JP21K18600, JP22H04933, JP22H01168, JP23H01124, JP23K22439 and JP23K25821) from the Japan Society for the Promotion of Science, by research support funding from the Kyoto University Foundation, by ISHIZUE 2024 of Kyoto University Research Development Program, by Murata Science and Education Foundation, and by the JGC-S Scholarship Foundation.
Liquid helium is supplied by the Low Temperature and Materials Sciences Division, Agency for Health, Safety and Environment, Kyoto University.
\end{acknowledgment}
\bibliographystyle{25675}
\bibliography{25675}
\end{document}